\documentclass[preprint,showkeys,secnumarabic,showpacs,amsmath,amssymb]{revtex4}
\begin{document}
\date{\today}
\title{A 5D noncompact Kaluza -Klein cosmology in the presence of Null perfect fluid}

\author{Hossein Farajollahi}
\email{hosseinf@guilan.ac.ir} \affiliation{Department of Physics,
University of Guilan, Rasht, Iran}

\author{Hamed Amiri}
\email{amiri.hame@gmail.com}

\affiliation{Department of Physics, University of Guilan, Rasht,
Iran}

\date{\today}

\begin{abstract}

For the description of the early inflation, and acceleration expansion of
the Universe, compatible with observational data,
the $5D$ noncompact Kaluza--Klein cosmology is investigated.
It is proposed that the $5D$ space
 is filled with a null perfect fluid, resulting a perfect fluid in
  $4D$ universe, plus one along the fifth dimension. By analyzing the reduced field equations for flat FRW model, we show the
  early inflationary behavior and current acceleration of the universe.

\end{abstract}



\maketitle

\section{Introduction}

Recent astronomical observations of distant supernovae lightcurves,
\cite{Riess}, \cite{Astier}, indicate that at early time the universe has gone through
an inflationary epoch and currently is undergoing an accelerated
phase of its expansion. Within the framework of cosmology,
 these obserevations can be explained by dark energy, which
 occupies almost seventy percent of the content of our universe at
present \cite{Koivisto}. A simple explanation for dark energy is the
cosmological constant, which acts like a perfect fluid with an
equation of state, and the energy density is associated with quantum
vacuum \cite{Linder}.

On the other hand, the early time inflation and late time
acceleration of the Universe may be explained by the modification of
gravitation, such as extra dimensional theories of gravity which
recently have received much interest \cite{Wesson1}. Of particular interest is the
cosmology of $5D$ pure geometry in noncompact Kaluza–-Klein theory,
known as space-–time-–matter (STM) theory, in which one do not need to
insert any matter into the $5D$ manifold by hand provided an
appropriate definition is made for the energy-momentum tensor of
matter in terms of the extra part of the geometry \cite{pWesson}. As a result,
instead, the matter appears in four dimensions induced by the $5D$
vacuum theory. Physically, the picture behind this interpretation is
that curvature in $5D$ space induces effective properties of matter
in $4D$ space-time. This idea is a consequence of the Campbell's
theorem which states that any analytic N-dimensional Riemannian
manifold can be locally embedded in an (N+1)--dimensional Ricci flat
Riemannian manifold \cite{Liu}, \cite{Seahra}.

One may receive extra motivations to work on this challenging problem
by recognizing that the energy density of scalar fields may also
contribute to the early inflation and current accelerating expansion
of the universe \cite{Chiba}. So instead of pure $5D$ geometry in noncompact
Kaluza–-Klein theory, we examine the theory with matter fields. Based on integration of these theories to explain the early
inflation and current acceleration, Darabi has studied a
$5D$ universe consists of a $5D$ geometry subject
to a $5D$ energy momentum tensor in the framework of
noncompact Kaluza -Klein thoery \cite{Darabi}. In another attempt, in the present study, we investigate the non
vacuum $5D$ space--time subject to a $5D$ energy momentum tensor which
takes the form of a null perfect fluid in the above framework. Physically, a null fluid by itself describes either gravitational radiation,
or some kind of nongravitational radiation or a combination of these two. In here, the null fluid
induced from higher dimension, divides itself into a
dark energy part and a matter part where for the $5D$ velocity of the fluid we have
$\textbf{U.U}=0$. From the $FRW$ symmetries of the model, the
energy-momentum tensor then is associated with the sum of two
perfect fluids, one in $4D$ space--time and one along the fifth
dimension. We then address the early
inflation and current acceleration of the universe based on the above prescription.

\section{The Model}

We will consider a general five-dimensional manifold with
coordinates $x^A$ $(A = 1, 2, 3, 4, 5)$ and metric tensor
$g_{AB}(x^C)$. The $5D$ interval is then given by

\begin{equation}
dS^2=g_{AB}dx^A dx^B,
\end{equation}where for simplicity we take
\begin{eqnarray}
 \nonumber  g_{5\mu}&=&0, \\
\nonumber g_{55}&=&\epsilon{\phi^2(x^\mu)},\\
  g_{\mu\nu}&=&g_{\mu\nu}(x^\mu).
\end{eqnarray}
In here $\epsilon^2=1$ and the signature of the scaler part of the
metric is left general to allow timelike or spacelike signature for the
fifth dimension without loss of generality. The fifth dimensional independency of the scalar field and the 4D metric in our formalism does not contradict the non compactness of the theory . In the compact version of the theory the cylindrical condition imposed to justify the non observability of the 5th dimension. However, in here, the theory is noncompact but for simplicity we assume that the scalar field and the 4D metric is independent of this extra dimension. So it is a matter of how we justify the non observability and independency of the extra dimension.

The field equations for $5D$ non vacuum Einstein equations are,
\begin{equation}
 G_{AB}=8\pi G T_{AB},
\end{equation}
where $G_{AB}$ and $T_{AB}$ are the $5D$ Einstein tensor and
energy-momentum tensor respectively and the $5D$ Einstein tensor is
\begin{equation}
G_{AB}=R_{AB}-\frac{1}{2}g_{AB}R_{(5)}.
\end{equation}
By a minimal extension of general relativity, the five dimensional
Ricci tensor is defined in terms of the christoffel symbols exactly
as in four dimensions,
\begin{equation}
R_{AB}=\partial_C \Gamma^C_{AB}-\partial_B\Gamma^C_{AC}+\Gamma^C_{AB}
\Gamma^D_{CD}-\Gamma^C_{AD}\Gamma^D_{BC}. \label{Riccitensor}\\
\end{equation}
By a dimensional reduction and after canceling out derivatives with
respect to fifth coordinate $x^{5}$, the 4D part of the $5D$  Ricci
tensor is obtained by putting $A\rightarrow\mu$, $B\rightarrow\nu$
in equation (\ref{Riccitensor}) and expanding the summed terms on
the right hand side by letting $ C\rightarrow\lambda,4 $, etc. Therefore, we
have

\begin{eqnarray}
\hat{R}_{\mu\nu}&=&\partial_\lambda\Gamma^\lambda_{\mu\nu}+\partial_5\Gamma^5_{\mu\nu}
-\partial_\nu\Gamma^\lambda_{\mu\lambda}-\partial_\nu\Gamma^5_{\mu5}+
\Gamma^\lambda_{\mu\nu}\Gamma^\mu_{\lambda\mu}\nonumber\\
&+&\Gamma^\lambda_{\mu\nu}\Gamma^5_{\lambda5}
+\Gamma^5_{\mu\nu}\Gamma^D_{5D}
-\Gamma^\eta_{\mu\lambda}\Gamma^\lambda_{\nu\eta}-\Gamma^5_{\mu\lambda}\Gamma^\lambda_{\nu5}-
\Gamma^D_{\mu5}\Gamma^5_{\nu D},\label{Riccitensor4d}
\end{eqnarray}
where '$\hat{}$' denotes the $4D$ part of the $5D$ quantities. One finds
that the 4D Ricci tensor is a part of equation (\ref{Riccitensor4d})
that may be rewritten as
\begin{equation}
\hat{R}_{\mu\nu}=R_{\mu\nu}+\partial_{5}\Gamma^{5}_{\mu\nu}-
\partial_{\nu}\Gamma^{5}_{\mu\nu}+\Gamma^{\lambda}_{\mu\nu}\Gamma^{5}_{\lambda5}
+\Gamma^{5}_{\mu\nu}\Gamma^{D}_{5D}-\Gamma^{5}_{\mu\lambda}\Gamma^{\lambda}_{\nu5}
-\Gamma^{D}_{\mu5}\Gamma^{5}_{\nu D}.\label{Riccitensor4dd}
\end{equation}
After evaluating the Christoffel symbols in (\ref{Riccitensor4dd}),
we obtain,
\begin{equation}
\hat{R}_{\mu\nu}=R_{\mu\nu}-\frac{\nabla_{\mu}\nabla_{\nu}\phi}{\phi}.
\end{equation}
We also find the $55$ component of the Ricci tensor as
\begin{equation}
R_{55}=\partial_{\lambda}\Gamma^{\lambda}_{55}-\partial_{5}\Gamma^{\lambda}_{5\lambda}
+\Gamma^{\lambda}_{55}\Gamma^{\eta}_{\lambda\eta}+\Gamma^{5}_{55}\Gamma^{\eta}_{5\eta}
-\Gamma^{\lambda}_{5\eta}\Gamma^{\eta}_{5\lambda}-\Gamma^{5}_{5\eta}\Gamma^{\eta}_{55}.
\end{equation}
Evaluation of the corresponding Christoffel symbols leads to
\begin{equation}
R_{55}=-\epsilon\phi\Box\phi.
\end{equation}
We then find the $5D$ Ricci scalar $R_{(5)}$ as
\begin{eqnarray}
R_{(5)}&=&g^{AB}R_{AB}=\hat{g}^{\mu\nu}R_{\mu\nu}+g^{55}R_{55}\nonumber\\
       &=&g^{\mu\nu}(R_{\mu\nu}-\frac{\nabla_{\mu}\nabla_{\nu}\phi}{\phi})+
\frac{\epsilon}{\phi^{2}}(-\epsilon\phi\Box\phi)=R-\frac{2}{\phi}\Box\phi,
\end{eqnarray}
where the $\mu5$ terms vanish and $R$ is the $4D$ Ricci scalar.

We are now ready to construct the space-time and $55$ components of
the Einstein tensor. The space-time components of the Einstein
tensor after substituting $R_{\mu\nu}$ and $R_{(5)}$ are
\begin{equation}
\hat{G}_{\mu\nu}=\hat{R}_{\mu\nu}-\frac{1}{2}\hat{g}_{\mu\nu}R_{(5)}
=G_{\mu\nu}+\frac{1}{\phi}(g_{\mu\nu}\Box\phi-\nabla_{\mu}\nabla_{\nu}\phi).\label{ET}
\end{equation}
In the same way, the $55$ component after substituting $R_{55}$
 and $R_{(5)}$ is
\begin{equation}
G_{55}=R_{55}-\frac{1}{2}R_{(5)}=-\frac{1}{2}\epsilon R
\phi^{2}.\label{ET55}
\end{equation}
\section{The FRW model}

Consider that the $4D$ space--time metric has the FRW line element,
\begin{equation}
ds^2=dt^2-R(t)^2\left(\frac{dr^2}{1-kr^2}
+r^2(d\theta^{2}+\sin\theta^{2}d\phi^{2})\right)\label{FRW},
\end{equation}
where $k$ takes the values $+1$,$0$,$-1$ according to a close, flat
or open universe, respectively.

We now assume that the energy momentum tensor in $5D$ takes the form of a
null anisotropic perfect fluid, where for the velocity of the fluid in $5D$ we have
$\textbf{U.U}=0$, where $U$ is velocity in terms of 5D proper time. From the symmetries of the metric (\ref{FRW}), we also assume that the
energy-momentum tensor is associated with the sum of two
perfect fluids, one in $4D$ space--time and one along the fifth
dimension. The energy-momentum tensor in 4D space time is
\begin{equation}
  T_{\mu\nu} = (\rho+p)u_{\mu}u_{\nu}-pg_{\mu\nu},\label{em} \\
\end{equation}
where $\rho$ and $p$ are the energy density and pressure in $4D$
space--time. We also obtain the energy momentum tensor along the fifth
dimension after writing the 5th component of the proper velocity in terms of 4D proper time:
\begin{equation}
  T_{55} =
  \frac{\overline{\rho}+\overline{p}}{1-\epsilon \phi^2}u_{5}u_{5}-\overline{p}g_{55},\label{em55}
\end{equation}
where $\overline{\rho}$ and $\overline{p}$ are the energy density
and pressure along the fifth dimension. In both (\ref{em}) and (\ref{em55}) velocities are in terms of 4D proper time

Using energy- momentum tensors (\ref{em}), (\ref{em55}), and the 4D
and fifth components of Einstein tensor, (\ref{ET}) and
(\ref{ET55}), the full non-vacuum $4D$ and the fifth dimensional Einstein
equations can be obtained as
\begin{eqnarray}
G_{\mu\nu}&=&8\pi G
[(\rho+p)u_{\mu}u_{\nu}-pg_{\mu\nu}]+\frac{1}{\phi}[\nabla_{\mu}\nabla_{\nu}\phi-\Box\phi
g_{\mu\nu}],\label{einstein4D}\\
R&=&16\pi G [\frac{\overline{\rho}+\overline{p}}{1-\epsilon \phi^2}+\overline{p}]\label{einstein55}.
\end{eqnarray}
Taking the trace of equation (\ref{einstein4D}) and combining with equation (\ref{einstein55}), one obtains
\begin{equation}
\Box\phi=\frac{1}{3}\{8\pi G(\rho-3p)+16\pi
G(\frac{\overline{\rho}+\overline{p}}{1-\epsilon \phi^2}+\overline{p})\}\phi,\label{scalarfieldeq}
\end{equation}
where '$\Box$' is defined as usual in $4D$ by
$\Box\phi\equiv g^{\mu\nu}\nabla_{\mu}\nabla_{\nu}\phi$. For a nonvanishing $\phi$, from equation (\ref{scalarfieldeq})one may assume the following
 replacements
\begin{equation}
\frac{1}{\phi}\Box\phi=\frac{1}{3}[8\pi G(\rho-3p)+16\pi
G(\frac{\overline{\rho}+\overline{p}}{1-\epsilon \phi^2}+\overline{p})]\label{delamberian},
\end{equation}
\begin{equation}
\frac{1}{\phi}\nabla_{\mu}\nabla_{\nu}\phi=\frac{1}{3}[8\pi
G(\rho-3p)+16\pi G(\frac{\overline{\rho}+\overline{p}}{1-\epsilon \phi^2}+\overline{p})] u_{\mu}u_{\nu}.\label{phideriv}
\end{equation}

Substituting equations (\ref{delamberian}) and (\ref{phideriv}) into equation (\ref{einstein4D}) leads to
\begin{equation}
G_{\mu\nu}=8\pi G
[(\rho+\widetilde{p})u_{\mu}u_{\nu}-\widetilde{p}g_{\mu\nu}],\label{finalEtensor}
\end{equation}
where
\begin{equation}
\widetilde{p}=\frac{1}{3}(\rho+2(\frac{\overline{\rho}+\overline{p}}{1-\epsilon \phi^2}+\overline{p})),\label{finalpressure}
\end{equation}
is the equivalent pressure in $4D$. The right hand side of equation (\ref{finalEtensor}) describes a perfect fluid
 with density $\rho$ and pressure $\widetilde{p}$. It is interesting that the contributions
 of the scalar field at higher dimension cancels out exactly
the physics of pressure $p$ in $4D$ and instead
substitute the pressure $\widetilde{p}$ given by equation
(\ref{finalpressure}) in terms of the matter density , dark
energy density and dark pressure density.

The field equations (\ref{finalEtensor}) for the metric (\ref{FRW})
lead to two independent equations,
\begin{equation}
3\frac{\dot{a}^{2}+k}{a^{2}}=8\pi G \rho\,\label{frweq},
\end{equation}
\begin{equation}
\frac{2a\ddot{a}+\dot{a}^{2}+k  }{a^{2}}=-8\pi G \widetilde{p}.\label{frwacc}
\end{equation}
Differentiating equation (\ref{frweq}) and combining with
acceleration equation (\ref{frwacc}) leads to the conservation
equation,
\begin{equation}
\frac{d}{dt}(\rho a^{3})+\widetilde{p}\frac{d}{dt}(a^{3})=0.\label{conserv}
\end{equation}
Using equation (\ref{frweq}), the acceleration equation
(\ref{frwacc}) can be rewritten as
\begin{equation}
\frac{\ddot{a}}{a}=-\frac{4\pi
G}{3}(\rho+3\widetilde{p})=-\frac{8\pi G
}{3}(\rho+\bar{\rho}+2\overline{p}).\label{acc}
\end{equation}
From equation (18) we obtain
\begin{equation}\label{extra}
-\frac{6(\kappa+\dot{a}^{2}+\ddot{a}a)}{a^2}=16\pi
G(\bar{\rho}+2\bar{p}).
\end{equation}
Using power law behaviors for scale factor, dark pressure and dark
density,
\begin{equation}
a(t)=a_{0}t^{\alpha},\ \ \  \bar{p}(t)=\bar{p}_{0}t^{\beta}, \ \ \
\bar{\rho}(t)=\bar{\rho}_{0}t^{\eta}\nonumber
\end{equation}
in equation (\ref{extra}), one can easily find that $\beta=\eta=-2$.

We also find that for a time dependent scaler field, the equation
(\ref{scalarfieldeq}) becomes
\begin{equation}
\ddot{\phi}+3\frac{\dot{a}}{a}\dot{\phi}=\frac{1}{3}\{8\pi
G(\rho-3p)+16\pi G(\overline{\rho}+2\overline{p})\}\phi.
\label{scalarfieldeq1}
\end{equation}
Using power law behavior for scalar field and matter density,
\begin{equation}
\phi(t)=\phi_{0}t^{\gamma} \ \ \rho(t)=\rho_{0}t^{\delta} \
(\rho_{0}>0)\nonumber
\end{equation}
with the equation of state for matter pressure, $p=\omega \rho$, and
dark pressure, $\bar{p}=\Omega \bar{\rho}$, for a flat universe
($k=0$), we rewrite the acceleration equation, (\ref{acc}), scaler
field equation, (\ref{scalarfieldeq1}), and conservation equation,
(\ref{conserv}), respectively in the following form
\begin{eqnarray}
&&\alpha(\alpha-1)+\frac{8\pi G}{3}(\rho_{0}+\bar{\rho_{0}}(1+2\Omega))=0, \label{acc2}\\
&&\gamma(\gamma-1)+3\alpha\gamma-\frac{8\pi G}{3}(\rho_{0}(1-3\omega)+2\bar{\rho_{0}}(1+2\Omega))=0,
 \label{scalarfieldeq2} \\
&&2\rho_{0}(2\alpha-1)+2\bar{\rho_{0}}\alpha(1+2\Omega)=0\label{conserv2},
\end{eqnarray}
where due to consistency in the conservation equation we have $\delta=-2$.

If we also define a new equation of state, $\widetilde{p}=\Gamma
\rho$, then, from equation (\ref{finalpressure}), we obtain
\begin{equation}
\Gamma=\frac{2-3\alpha}{3\alpha}.\label{ratio}
\end{equation}
One can simply check that in radiation dominant era where
$\alpha=1/2$, we have $\Gamma=1/3$, and in matter dominant era where
$\alpha=2/3$, we have $\Gamma=0$, as expected. One also find that in
early inflationary era where $\alpha >>1$, we have
$\Gamma\simeq -1$ and in late time acceleration that $\alpha >1$ we
have $\Gamma<-1/3$. Therefore, the new equivalent equation of state, $\widetilde{p}=\Gamma
\rho$ instead of $p=\omega \rho$ or $\bar{p}=\Omega \bar{\rho}$, satisfies the physical
conditions on different epochs.

From acceleration equation (\ref{acc}), for an accelerating
universe, and with the help of equation (\ref{conserv2}) we require
that $\Omega<-1/2$ and $\alpha>1$ which accounts for a negative dark
energy and accelerating universe.

In radiation dominant era where $\alpha=1/2$, from
equations (\ref{scalarfieldeq2}) and (\ref{conserv2}) one finds that
\begin{equation}
-1/2<\gamma<0.
\end{equation}
Also, in matter dominant era where $\alpha=2/3$, we
find that
\begin{equation}
-1<\gamma<0.
\end{equation}
As it can be seen, the
larger values of $\alpha$ lead to more negative value of $\gamma$
which means that as the universe evolves in time from radiation to
matter state the scalar field is more suppressed. One also
  find in acceleration equation (\ref{acc}) that for  larger values of $|\Omega|$, the value
  of $\alpha$ is larger and larger which means that the more negative pressure we have,
  the more accelerated universe expected.

From equation (\ref{conserv2}), we define the ratio of dark energy
density to matter density as
\begin{equation}
r=\frac{\bar{\rho_{0}}}{\rho_{0}}=\frac{1-2\alpha}{\alpha(1+2\Omega)}.\label{ratio}
\end{equation}

From equation (\ref{ratio}) in radiation dominated era where
$\alpha=1/2$, we obtain $r=0$ or the energy density $\bar{\rho}=0$
which means that the time evolution for the scale factor is not
affected by the dark energy. In addition, in the matter dominated era, where
$\alpha=2/3$, in order to have a positive dark energy density, we
find that $\Omega<-1/2$. On the other hand, in the early universe
inflationary era for $\alpha >> 22$, we have $\Omega
\simeq-1/2-1/(4r)$, and for an accelerating universe in the late
time where $\bar{\rho}>\rho$ and $\alpha>1$, we have
$\Omega<-1/2-1/(2\alpha)$.

\section{ Conclusion}
In this paper, we present a $5D$ universe subject to a
$5D$ energy-momentum tensor in the framework of
noncompact Kaluza-Klein theory. For simplicity we assume that the
metric of the space does not depend explicitly on the extra coordinate and also
$g_{\mu 5}=0$. We also represent the energy momentum tensor in $5D$ space--time as a null perfect fluid.
 The geometry of the $4D$ space--time is taken to
be FRW subject to the conventional perfect fluid
with density $\rho$ and pressure $p$ while the extra dimensional
part endowed by a scalar field is subject to another perfect fluid
with dark density density $\overline{\rho}$ and dark energy pressure
$\overline{p}$. By writing down the reduced $4D$ and extra-dimensional
components of $5D$ Einstein equations we find that the $4D$
universe corresponds to the vacuum states of the scalar field. It
turned out that the contribution of the non--vacuum states of the
scalar field to the $4D$ cosmology cancels out exactly the physics of
pressure $p$ and instead require a new equivalent pressure
$\widetilde{p}$ in $4D$ that leads to non-zero
late acceleration of the universe and also satisfies some of the observational
 constraints including equation of state constants in matter and radiation dominated era and early inflation.

\end{document}